%% file: paper.tex
\PassOptionsToPackage{table}{xcolor}
\documentclass[sigconf,10pt,screen]{acmart}

\usepackage{amsmath,amsfonts}
\usepackage{algorithmic}
\usepackage{graphicx}
\usepackage{textcomp}
\usepackage{url}
\usepackage{soul}
\usepackage{tikz}

\def\BibTeX{{\rm B\kern-.05em{\sc i\kern-.025em b}\kern-.08em
    T\kern-.1667em\lower.7ex\hbox{E}\kern-.125emX}}

\usepackage{listings}
\usepackage{xspace}
\usepackage{todonotes}
\usepackage{ctable}
\usepackage{tablefootnote}
\usepackage{subcaption}
\usepackage{dirtytalk}

\usepackage{makecell}

\usepackage{booktabs}
\usepackage{xparse}

\NewDocumentCommand{\rot}{O{45} O{1em} m}{\makebox[#2][l]{\rotatebox{#1}{#3}}}

\usepackage{mathrsfs}
\usepackage{mathtools}
\usepackage{phoenician}

\usepackage{annotate-equations}

\newcommand{\pp}{\reflectbox{$\mathrm{P}$}\mkern-7mu\mathrm{P}}

\newcommand{\ppm}{$\pp$\xspace}

\newcommand{\zaratan}{Zaratan}

\definecolor{obshl}{RGB}{63, 90, 126}
\definecolor{obsbg}{RGB}{228, 244, 247}

\newcounter{obsnum}

\newcommand{\observe}[2]{%
  \refstepcounter{obsnum}\label{#1}%
  \setlength{\fboxrule}{0mm}%
  \setlength{\fboxsep}{2mm}
  \vspace{0.35em}%
  \noindent
  \colorbox{obshl}{%
    \parbox[0mm]{\columnwidth - 2\fboxsep}{%
      \textbf{\textcolor{white}{Observation \theobsnum}}
  }}%
  \newline%
  \fcolorbox{obshl}{obsbg}{%
    \parbox[0mm]{\columnwidth - 2\fboxsep - 2\fboxrule - 0.9mm}{%
      #2%
    }
  }
}

\makeatletter
\g@addto@macro\UrlSpecials{\do\!{\newline}}
\makeatother

\newcommand{\tweakedsim}{\raise.17ex\hbox{$\scriptstyle\mathtt{\sim}$}}

\definecolor{codegreen}{rgb}{0,0.6,0}
\definecolor{codegray}{rgb}{0.5,0.5,0.5}
\definecolor{codepurple}{rgb}{0.58,0,0.82}
\definecolor{backcolour}{rgb}{0.95,0.95,0.92}

\definecolor{boxyell}{HTML}{FFCC00}
\newcommand\ybx[1]{\cellcolor{boxyell}\bfseries #1}
\newcommand\gbx[1]{\cellcolor{codegreen}\bfseries \textcolor{white}{#1}}

\colorlet{soulgreen}{codegreen!20}
\colorlet{soulpurple}{codepurple!20}
\colorlet{soulyell}{boxyell!20}

\DeclareRobustCommand{\highlight}[2]{{\sethlcolor{#1}\hl{#2}}}
\soulregister{\highlight}{2}

\lstdefinestyle{mystyle}{
    backgroundcolor=\color{backcolour},
    commentstyle=\color{codegreen},
    keywordstyle=\color{magenta},
    numberstyle=\tiny\color{codegray},
    stringstyle=\color{codepurple},
    basicstyle=\ttfamily\footnotesize,
    breakatwhitespace=false,
    breaklines=true,
    captionpos=b,
    keepspaces=true,
    numbers=left,
    numbersep=5pt,
    showspaces=false,
    showstringspaces=false,
    showtabs=false,
    tabsize=2
}

\newcommand{\umdcs}{\affiliation{%
  \institution{Department of Computer Science, University of Maryland}
  \city{College Park}
  \state{Maryland}
  \country{USA} 
}}
\newcommand{\casc}{\affiliation{%
  \institution{Lawrence Livermore National Laboratory}
  \city{Livermore}
  \state{California}
  \country{USA}
}}

\copyrightyear{2025}
\acmYear{2025}
\setcopyright{cc}
\setcctype{by-nc-sa}
\acmConference[ICS '25]{2025 International Conference on Supercomputing}{June 8--11, 2025}{Salt Lake City, UT, USA}
\acmBooktitle{2025 International Conference on Supercomputing (ICS '25), June 8--11, 2025, Salt Lake City, UT, USA}
\acmDOI{10.1145/3721145.3730423}
\acmISBN{979-8-4007-1537-2/2025/06}

\begin{document}
\tolerance=5000

\title{Taking GPU Programming Models to Task for Performance Portability}

\author[J.~H.~Davis]{Joshua H.~Davis}
\umdcs
\email{jhdavis@umd.edu}
\author[P.~Sivaraman]{Pranav Sivaraman}
\umdcs
\email{psivaraman@umd.edu}
\author[J.~Kitson]{Joy Kitson}
\umdcs
\email{jkitson@umd.edu}
\author[K.~Parasyris]{Konstantinos Parasyris}
\casc
\email{parasyris1@llnl.gov}
\author[H.~Menon]{Harshitha Menon}
\casc
\email{gopalakrishn1@llnl.gov}
\author[I.~Minn]{Isaac Minn}
\umdcs
\email{iminn@umd.edu}
\author[G.~Georgakoudis]{Giorgis Georgakoudis}
\casc
\email{georgakoudis1@llnl.gov}
\author[A.~Bhatele]{Abhinav Bhatele}
\umdcs
\email{bhatele@cs.umd.edu}

\def \authors{Joshua H. Davis, Pranav Sivaraman, Joy Kitson, Konstantinos Parasyris, Harshitha Menon, Isaac Minn, Giorgis Georgakoudis, Abhinav Bhatele}

\begin{abstract}
  \input{abstract}
\end{abstract}




\maketitle

\section{Introduction}
\label{sec:intro}
\input{intro}

\section{Background: Portable Programming Models}
\label{sec:bg}
\input{bg}

\section{Related Work}
\label{sec:related}
\input{related}

\section{Methodology for Evaluating Performance Portability on GPU Platforms}
\label{sec:method}
\input{method}

\section{Porting to Unsupported Programming Models}
\label{sec:porting}
\input{porting}

\section{Experimental Setup}
\label{sec:setup}
\input{setup}

\section{Results and Discussion}
\label{sec:results}
\input{results}

\section{Conclusion}
\label{sec:conclusion}
\input{conc}

\begin{acks}
\input{ack}
\end{acks}

\bibliographystyle{ACM-Reference-Format}
\bibliography{./bib/cite,./bib/pssg}

\end{document}

%% file: abstract.tex
Portability is critical to ensuring high productivity in developing and
maintaining scientific software as the diversity in on-node hardware
architectures increases. While several programming models provide portability
for diverse GPU systems, they don't make any guarantees about performance
portability. In this work, we explore several programming models -- CUDA, HIP,
Kokkos, RAJA, OpenMP, OpenACC, and SYCL, to assess the consistency of their
performance across NVIDIA and AMD GPUs. We use five proxy applications from
different scientific domains, create implementations where missing, and use
them to present a comprehensive comparative evaluation of the performance
portability of these programming models. We provide a Spack scripting-based
methodology to ensure reproducibility of experiments conducted in this work.
Finally, we analyze the reasons for why some programming models underperform in
certain scenarios and in some cases, present performance optimizations to the
proxy applications.

%% file: intro.tex
Heterogeneous CPU-GPU architectures have come to dominate the design of high
performance computing (HPC) systems. Nine of the top ten systems in the November
2024 TOP500 list, and \tweakedsim 42\% of the systems on the complete list,
employ co-processors or accelerators~\cite{top500nov2024}. Further, a diverse
set of specific architectures are in use, supplied by a range of vendors, as the
current top ten includes GPUs from AMD, NVIDIA, and Intel. A similarly diverse
range of programming models have emerged, which all aim to allow application
developers to write their code once and run it on any system. Programming models
such as OpenMP~\cite{OpenMP4}, RAJA~\cite{RAJA}, and
Kokkos~\cite{kokkos:tpds2022} act as \textit{portability layers}, bridging the
gap between high-level implementation of an algorithm and low-level execution on
a given target architecture. Yet running scientific applications efficiently on
HPC systems requires more than just \textit{functional} portability, which
refers to program correctness. Codes must also perform well on a range of target
systems, ideally without incurring the technical debt of system-specific
implementations. This is often referred to as \textit{performance} portability.

Application developers would benefit from a deeper understanding of the
performance portability provided by different programming models on modern GPU
systems before porting their application to a particular model. Choosing a
programming model for porting a CPU-only application to GPUs is a major
commitment, requiring significant time for developer training and programming.
If a programming model delivers unacceptable performance, then that investment
is wasted.

Nevertheless, each programming model's effectiveness at enabling performance
portability, as well as the definition of performance portability itself, remain
open questions. Although developers' experiences comparing the performance
portability of several models on a single application are valuable, we have
observed that open-source applications or even proxy applications implemented in
a several different programming models are uncommon and difficult to find.
Further, a single smaller application or benchmark implemented in most
programming models is unlikely to be representative of the diverse and complex
production applications typically run on HPC systems. Finally, conducting
exhaustive combinatorial studies of programming model, compiler, system, and
application combinations is a significant undertaking, as each programming model
usually requires unique combinations of compilers flags and libraries for any
given system.

In this paper, we provide a comprehensive empirical study of the performance
portability of several programming models on GPU-based leadership-class
supercomputers. We use a variety of proxy applications that are representative
of production codes, and using them, we enable realistic comparisons of the
performance portability of GPU kernels written in several programming models
across different architectures. We study five proxy applications from different
scientific domains, create implementations where missing, and comprehensively
evaluate differences between these programming models.

We present a Spack-based~\cite{gamblin:sc15} environment and scripting system to
significantly lower the barrier for performance portability studies.
This system encapsulates our methodology for systematically building, running
and benchmarking a suite of applications in several programming models, in a
manner which can be adapted for future studies. Our comparative evaluation of
model performance includes specific insights into why certain
programming models perform well or poorly for particular applications on
different target systems. To our knowledge, this is one of the most
comprehensive performance portability studies to date, in terms of the breadth
of programming models and applications studied and the detail provided in the
analysis of results.

To summarize, our contributions include the following:
\begin{itemize}
  \item We evaluate the performance portability enabled by seven different
programming models using a diverse set of five proxy applications benchmarked
across NVIDIA and AMD GPUs on production HPC systems.
  \item We create several additional implementations of existing proxy
applications in previously unsupported programming models to ensure full coverage of programming
models across applications.
  \item We describe a methodology employing Spack scripting and environment
tools~\cite{gamblin:sc15} to easily manage the process of building and running
all $7 \times 5 = 35$ versions across five supercomputing systems, each with
unique software stacks. We open-source these recipes for the community in order to
substantially reduce the effort required to reproduce or extend our study.
  \item We conduct a thorough analysis of the reasons for key outliers in the
performance portability cases studied, and describe and test optimizations that
improve performance portability in some cases.
\end{itemize}

%% file: bg.tex
In this section, we provide relevant background information on the various
programming models we evaluate. HIP and CUDA act as our baselines in
this study, as they are the native models for AMD and NVIDIA
devices, respectively. Below, we describe the key attributes of each
category of programming model. All programming models used in this study
support both NVIDIA and AMD devices except CUDA.

\vspace{0.08in}
\noindent \textbf{Language extensions:}
SYCL, HIP, and CUDA are language extensions, which add features to the base
language (C++, C, and/or Fortran) for programming GPUs. SYCL and HIP are open
standards, while CUDA is proprietary. The language extensions we consider are
more verbose than the other programming models. Users call runtime functions to
manage memory and write functions that they then invoke as kernels to offload
execution. SYCL provides multiple methods of memory management, including
the \emph{explicit USM (unified shared memory)} API, which uses CUDA or HIP
style runtime calls to move and allocate data, or the
\emph{buffer/accessor} API, which is more implicit, allowing the compiler and
runtime to schedule data movement but not allowing explicit access to valid
device pointers.

\vspace{0.08in}
\noindent \textbf{C++ abstraction libraries:}
Kokkos and RAJA are C++ abstraction libraries. These are template-based C++
libraries that provide high-level functions and data types. Users write their
code directly employing these data types and typically structure GPU code as
lambdas to pass into library function calls. The library translates the user
code to a device backend such as CUDA, HIP, or OpenMP at compile-time or
runtime. Note that Kokkos provides both memory and compute abstractions, while
RAJA provides compute abstractions and users must employ the related Umpire or
CHAI libraries to abstract memory management.

\vspace{0.08in}
\noindent \textbf{Directive-based models:}
OpenMP and OpenACC are directive-based models. They provide compiler directives,
or \emph{pragmas}, to parallelize or offload code. They are typically standard
specifications implemented by a compiler front-end and a runtime library to
implement parallel or offloaded execution that abstracts the underlying hardware
architecture. Directive-based models are usually less verbose and less
intrusive, as users can often annotate existing code with minimal refactoring.
This facilitates incremental development. These models provide clauses and
standalone directives to schedule data movement, which is carried out by
the compiler and device runtime.

%% file: related.tex
Several studies on programming language extensions, models, and libraries have
been designed to assist developers achieve performance
portability~\cite{OpenMP4,sabne2015evaluating,ben2019stateful,RAJA,kokkos:tpds2022}.
Additionally, several studies have assessed the portability of certain
frameworks. We categorize the related work on empirical performance portability
studies into three groups: metric studies, application or programming model
studies, and broader studies that are not scoped to a particular model or app.
Below, we provide an overview of recent work in each category.

\vspace{0.08in}
\noindent \textbf{Studies of performance portability metrics:}
Pennycook et al. propose the metric \ppm for performance portability, defining
it as the harmonic mean of the performance efficiencies of an application across
different systems~\cite{pennycook2016metric, pennycook2019implications,
sewall2020interpreting, pennycook2021navigating, pennycook2021revisiting}.
Daniel et al. propose an alternative metric, $P_D$, which accounts for problem
size~\cite{daniel2019applying}, and Marowka compares \ppm with
$\overline{\pp}$, a similar metric that uses the arithmetic mean instead of
the harmonic mean~\cite{marowka2023comparison,marowka2021toward}.

\vspace{0.08in}
\noindent \textbf{Studies involving individual application categories or programming models:}
A number of studies evaluate performance portability in specific applications
with multiple programming models or a single programming model
model~\cite{martineau2017assessing, reguly2020productivity,
reguly2019performance, sedova2018high, boehm2018evaluating, dufek2021case,
deakin2018evaluating, artigues2020evaluation, rangel2023performance,
gayatri2019case, sabne2015evaluating, brunst2022first,kuncham2021performance,
karlin:ipdps2013}. For instance, Dufek et al. compare Kokkos and SYCL for the
Milc-Dslash benchmark~\cite{dufek2021case}, while Rangel et al. examine the
portability of CRK-HACC in SYCL~\cite{rangel2023performance}. Other studies
investigate performance portability across applications using specific
programming models. Brunst et al. benchmark the 2021 SPEChpc suite,
which contains nine mini-applications in OpenMP and OpenACC, on Intel CPUs and
NVIDIA and AMD GPUs~\cite{brunst2022first}. Kuncham et al. evaluate the
relative performance of SYCL and CUDA on the NVIDIA V100 using BabelStream,
Mixbench, and Tiled Matrix-Multiplication~\cite{kuncham2021performance}.

While these studies provide useful information to developers working on similar
applications or those interested in specific programming models, making more
general statements about programming models themselves requires a broader
evaluation of a diverse set of case studies.

\begin{table*}[t]
  \centering
  \caption{Summary of proxy applications and benchmarks used in this study
    along with which of their ports the authors changed. Here, E = already exists,
    {\bfseries \highlight{boxyell}{ M }} = modified by us,
    \textcolor{white}{\bfseries \highlight{codegreen}{ C }} = created by us.}
{\small
  \begin{tabular}{lccc|cccccccc}
    \toprule
    Proxy Application & Scientific Domain  & Method(s)                                     & Suite   & \rot{CUDA} & \rot{HIP} & \rot{SYCL} & \rot{Kokkos} & \rot{RAJA} & \rot{OpenMP} & \rot{OpenACC} & \rot{Spack Pkg.} \\ \midrule
    BabelStream       & N/A                & Bandwidth benchmark                           & N/A     & E          & E         & E          & E            & \ybx{M}    & E            & E             & \ybx{M} \\
    XSBench           & Nuclear physics    & Monte Carlo                                   & ECP     & E          & E         & \ybx{M}    & \gbx{C}      & \gbx{C}    & E            & \gbx{C}       & \ybx{M} \\
    CloverLeaf        & Hydrodynamics      & Structured grid                               & Mantevo & E          & E         & \ybx{M}    & E            & \gbx{C}    & E            & \gbx{C}       & \ybx{M} \\
    su3\_bench        & Particle physics   & \makecell{Structured grid,\\ dense lin.~alg.} & NERSC   & E          & E         & E          & E            & \gbx{C}    & E            & E             & \gbx{C} \\
    miniBUDE          & Molecular dynamics & N-body                                        & N/A     & E          & E         & \ybx{M}    & E            & \ybx{M}    & E            & E             & \gbx{C} \\
    \bottomrule
  \end{tabular}
}
  \label{tab:development}
\end{table*}

\vspace{0.08in}
\noindent \textbf{Broader performance portability studies:}
Deakin et al.~present performance portability studies of five programming models
across a wide range of hardware architectures, using BabelStream, TeaLeaf,
CloverLeaf, Neutral, and MiniFMM~\cite{deakin2019performance,
deakin2020tracking}. More recent papers by Deakin et al. focus on more specific
problems such as reductions and GPU to CPU
portability~\cite{deakin2021analyzing, deakin2022heterogeneous}. Lin et al.
evaluate implementations of C++17 StdPar against five models on AMD
devices~\cite{lin2024preliminary}. While these studies provide performance
portability comparisons across systems, applications, and models, they do not
include RAJA and sometimes omit HIP and OpenACC. Furthermore, they do not
provide extensive analysis of the reasons for performance differences between
programming models or ways to address differences.

Several other studies are similar in scope but different in focus. Kwack et al.
evaluate portability development experiences for three full applications and
three proxy applications across GPUs from multiple
vendors~\cite{kwack2021evaluation}. Harrell et al. study performance portability
alongside developer productivity~\cite{harrell2018effective}. However, in these
studies each application is only ported to a single portable programming model.
This makes it difficult to draw conclusions about each programming model's
relative suitability to particular applications. Koskela et al. provide six
principles for reproducible portability benchmarking, along with a demonstration
of these principles in a Spack+Reframe CI infrastructure for a study of
BabelStream on some CPU architectures and an NVIDIA
V100~\cite{koskela2023principles}. Other studies uniformly fail to follow
these principles, making reproducing them an arduous task.

%% file: method.tex
In this section, we describe our approach to comprehensively compare programming
models that provide portability on GPU systems. We also justify for our choices
of programming models, proxy applications, systems, and metrics.

\subsection{Choice of programming models}

Our goal in this work is to empirically compare the performance portability
provided by popular programming models. In Section~\ref{sec:bg}, we describe
three categories of programming models with a few examples in each category. We
identify those representative models by surveying a broad range of proxy
applications in order to determine how common existing implementations in each
model are. We survey a variety of sources for proxy applications, including the
ECP Proxy Apps suite~\cite{ecp-proxy-apps}, the NERSC Proxy
suite~\cite{nersc-proxy-apps} and the Mantevo Applications
Suite~\cite{heroux2013mantevo}. Armed with that knowledge, we decide to focus on
CUDA, HIP, SYCL, Kokkos, RAJA, OpenACC, and OpenMP, as they are the most popular
models found in the proxy applications we surveyed. Together, these models cover
the three categories of models mentioned earlier.

\subsection{Choice of proxy applications}

Based on the survey of proxy applications mentioned above, we identify five
applications that represent the range of typical scientific computing workloads
on GPU clusters. These include a pure memory bandwidth benchmark as well as four
other proxy applications. They range from highly compute-intensive (miniBUDE) to
highly memory-intensive (BabelStream), and also include one representative from
each of the three large proxy application suites we surveyed. The scientific
domains covered by them include hydrodynamics (CloverLeaf), molecular dynamics
(miniBUDE), nuclear physics (XSBench), and particle physics (su3\_bench), and
computational methods include structured grid (CloverLeaf and su3\_bench), dense
linear algebra (su3\_bench), n-body (miniBUDE) and Monte Carlo (XSBench)
methods. The scientific domains represented cover three of the four most common
disciplines found in INCITE awardees of the last three years — physics,
engineering, and biology.\footnote{\url{https://doeleadershipcomputing.org/awardees/}}

CloverLeaf, miniBUDE, and XSBench are missing implementations in some
programming models compared in this work. So, we develop these missing
implementations to obtain full coverage of the space of application and model
combinations. Table~\ref{tab:development} summarizes the key details of each
proxy application, and identifies the implementations that are either created
or modified by us for this study. Our modifications consist of small changes to
the memory management library or style to ensure portability and consistency of
gathering execution times across implementations. Below, we describe
the five proxy applications used in this study:

\vspace{0.08in}
\noindent \textbf{BabelStream} is a memory bandwidth benchmark with five
kernels: {\tt copy, add, mul, triad,} and {\tt dot}~\cite{deakin2018evaluating}.
The dot kernel includes a reduction operation, known to be a challenging
operation for some programming models~\cite{davis2021performance}.

\vspace{0.08in}
\noindent \textbf{XSBench}~\cite{tramm2014xsbench} is a proxy for OpenMC, a
Monte Carlo transport code~\cite{romano2015openmc}. XSBench runs one kernel,
OpenMC's macroscopic cross-section lookup kernel, with a large number of
lookups. We use the event-based transport method with a hash-based grid as it is
preferred for GPUs.

\vspace{0.08in}
\noindent \textbf{CloverLeaf} is a 2D structured compressible Euler equation
solver, with 14 kernels~\cite{herdman2012accelerating}. The \verb|advec_mom|,
\verb|advec_cell|, \verb|PdV|, and \verb|calc_dt| kernels are typically the
most time-intensive, and \verb|calc_dt| contains a reduction.

\vspace{0.08in}
\noindent \textbf{su3\_bench}~\cite{doerfler2020su3_bench} is a proxy
application for MILC, a lattice quantum chromodynamics
code~\cite{bernard1991studying}. It implements the SU(3) matrix-matrix multiply
routine in its lone kernel.

\vspace{0.08in}
\noindent \textbf{miniBUDE} is a proxy for Bristol University Docking Engine
(BUDE), a molecular dynamics code which simulates molecular docking for drug
discovery~\cite{mcintosh2015high}. miniBUDE computes the energy field for
one configuration of a protein repeatedly.

\subsection{Choice of systems}

Evaluating performance portability requires selecting a range of systems with
diverse architectures. One of the main goals of this study is to evaluate
performance portability on production GPU-based supercomputers, given the rising
prominence of GPUs in new systems~\cite{top500jun2024}. We select five different
supercomputers for our experiments (architectural details in Table~\ref{tab:systems}): Summit and Frontier at ORNL, Perlmutter at
NERSC, Corona at LLNL, and \zaratan{} at the University of Maryland (UMD). These systems cover the
majority of the GPU architectures in the top ten systems. Frontier and Summit
are in the top ten, and Perlmutter is in the top fifteen. We include Corona (AMD
MI50) and \zaratan{} (NVIDIA H100) for context with older AMD and newer NVIDIA
hardware, respectively. For Frontier's MI250X GPUs, we run on one Graphics
Compute Die (GCD) which is an independent unit of allocation.

\begin{table}[h]
  \centering
  \caption{Architectural details of the GPUs in each system used in this paper.
    Flop/s and bandwidth values are theoretical peaks provided by device
    manufacturers.}
{\small
  \begin{tabular}{llrrr}
    \toprule
    System                                       & GPU Model   & \makecell[r]{Peak\\ Tflop/s*}  & \makecell[r]{DRAM\\ bandwidth} & \makecell[r]{DRAM\\ size} \\ \midrule
    \makecell[l]{Summit$^{\dagger}$\\ (ORNL)}      & NVIDIA V100 & 14.0/7.0                       & 900  GB/s                      & 32 GB \\
    \makecell[l]{Perlmutter\\ (LBL)}             & NVIDIA A100 & 9.5/9.7                        & 1555 GB/s                      & 40 GB \\
    \makecell[l]{\zaratan{}\\ (UMD)}             & NVIDIA H100 & 7.0/34.0                       & 3350 GB/s                      & 80 GB \\
    \makecell[l]{Corona\\ (LLNL)}                & AMD MI50    & 3.3/6.6                        & 1000 GB/s                      & 32 GB \\
    \makecell[l]{Frontier$^{\ddagger}$\\ (ORNL)}   & AMD MI250X  & 23.9/23.9                      & 1600 GB/s                      & 64 GB \\
    \bottomrule
  \end{tabular}
}
  \vspace{1mm}
  \flushleft
  \footnotesize{* Single-precision/double-precision.} \\
  \footnotesize{$^{\dagger}$ We use the high-memory GPUs on Summit.} \\
  \footnotesize{$^{\ddagger}$ Details for a single GCD of one MI250X.}
  \label{tab:systems}
\end{table}

\subsection{Measurement and evaluation strategy}
\label{sec:measurement}

In this study, we modify applications where needed to consider both the
efficiency of GPU kernel(s) and that of data movement between host and device
needed to run the application. However, as discussed in
Sec.~\ref{sec:results}, the impact of data movement on overall performance is
minimal for these applications and not presented in detail. We add a runtime
option to all the applications to specify a number of warmup iterations at the
start of the simulation which we exclude from timing. XSBench normally runs
only for only a single iteration, so we add a loop that repeatedly runs the
kernel a user-specified number of times to ensure consistency across
applications. As mentioned in Sec.~\ref{sec:setup}, variability across
runs is low, with runs of a given setup differing by at most 3.3\%.

Having determined how to consistently define performance for each application,
we can also derive additional higher-level metrics about performance portability
for each combination of application and programming model. In this work, we use
\textbf{\ppm with application efficiency} proposed by Pennycook et
al.~\cite{pennycook2016metric}. \ppm is defined, for some application
\highlight{soulyell}{$a$}, problem \highlight{soulpurple}{$p$}, set of systems
\highlight{soulgreen}{$H$}, and measure of application efficiency $e$, as:
\begin{equation*} \label{eq:ppm}
  \pp(\eqnmarkbox[boxyell]{a1}{a},\eqnmarkbox[codepurple]{p1}{p},
    \eqnmarkbox[codegreen]{h1}{H}) =
  \begin{cases}
    \dfrac{|\eqnmarkbox[codegreen]{h2}{H}|}{\sum_{i \in \eqnmarkbox[codegreen]{h3}{H}}
      \dfrac{1}{e_i(\eqnmarkbox[boxyell]{a2}{a},\eqnmarkbox[codepurple]{p2}{p})}}
      & \!\begin{aligned}
        \\
        & \text{if } i \text{ is\hspace{1em}supported }\\
        & \forall i \in \eqnmarkbox[codegreen]{h4}{H}
      \end{aligned}\\
    0 & \text{otherwise.}
  \end{cases}
  \vspace{1em}
\end{equation*}
\annotate[yshift=3em]{above, left}{h4,h2,h1}{set of systems}
\annotate[yshift=-1.4em]{below,right}{p1,p2}{problem}
\annotatetwo[yshift=-2.25em]{below}{a1}{a2}{application}

This is the harmonic mean of the efficiencies of an application running the same
input problem across a set of systems. The application efficiency $e_i(a, p)$ of
an application $a$ solving problem $p$ is the ratio $\frac{t_{min}}{t}$, where
$t$ is the runtime of $a$ solving $p$ on the particular hardware $i$, and
$t_{min}$ is the best observed runtime across all variants of $a$ solving $p$ on
$i$. \ppm ranges from 0 to 1, where 1.0 indicates the
application runs at the best observed performance on all systems.

\subsection{Automation and reproducibility strategy}
In our experiments, we ensure that compilers, dependency versions, and flags are
used consistently across applications and systems. We accomplish this with
Spack~\cite{gamblin:sc15}, a popular HPC package manager. We create a single
Spack environment file for each system which specifies the exact compiler,
application, and library dependency versions along with any needed flags. As
listed in Table~\ref{tab:development}, we have created or updated Spack package
files for each proxy app, and these updates will be provided to the community.
Our Spack environments for this project can be easily adapted to any new system,
allowing for easy reproduction of our experiments, and significantly reducing
the time-consuming effort of building every combination of application
and programming model.

We further employ Spack's Python scripting
tools\footnote{\url{https://spack-tutorial.readthedocs.io/en/latest/tutorial_spack_scripting.html}}
to develop robust automation for our experiments --- we can create jobs with a
single-line invocation leveraging Spack's spec syntax to adjust which
application, models, or compilers are used, and save profile data to disk to be
directly read by our plotting scripts. These scripts and environments will be
published to allow the community to use our portability study methodology. These
infrastructural contributions dramatically reduce the effort required to
reproduce our results and create new studies of portable programming models.

%% file: porting.tex
The proxy applications we choose have implementations in most of the evaluated
programming models. In these existing ports, we make minor modifications to
consistently align timing measurements across different programming models. We
also update the RAJA ports of BabelStream and miniBUDE to use Umpire for
portable memory allocations.

When creating new ports, we seek to apply the same level of effort for all of them in
order to avoid granting an unfair advantage to any particular implementation
arising from excess optimization. We spend similar amounts of time implementing
each new port, and keep the structure of the code between new and existing ports
as similar as possible. Further, we specifically do not tune kernel grid size,
block size, and shared memory per block. For programming models that require the
user to specify these values (CUDA, HIP, RAJA, SYCL), we use the default values
provided by the respective proxy application developers. For programming models
that can select their own default parameter values (OpenMP, OpenACC, Kokkos), we
allow the model to do so if compatible with the existing application code. Our
results reflect ``out of the box'' performance that a user would encounter with
minimal porting effort.

In the following subsections, we discuss our experiences working with the
programming models as applicable. Table~\ref{tab:development} summarizes our
development efforts. We plan to merge these contributions to their respective
upstream repositories.

\subsection{Porting to Kokkos}

Porting the XSBench code to Kokkos requires converting the existing {\tt for}
loop to be a lambda function passed into a {\tt Kokkos::parallel\_for} call and
converting the data structures to be used in Kokkos calls
to \verb|Kokkos:View|s. For example, XSBench's \verb|SimulationData| struct
contains several dynamic arrays, which need to be Views in order to work on the
GPU. In this situation, there are two options available to a developer: 1)
rewrite all of the application code to use Views from the beginning, including
any CPU-side setup or initialization; or 2) avoid rewriting the any setup code
by constructing Views out of pointers to any ordinary C++ arrays after
initialization but before copying them to the device and launching kernels.

We opted for the second of these methods to minimize changes to existing
application code. Listing~\ref{lst:kokkos} provides an example of this approach
as we implemented it. In summary, we construct an unmanaged View in
the \verb|HostSpace| called \verb|u_cocns| using the heap memory of
the \verb|SD.concs| array, construct a new View in the device space
called \verb|SD.d_concs|, and finally \verb|deep_copy| the unmanaged host View
to the new device View. While Kokkos requires developers to use its memory
abstraction, the View, in order to make use of its portable kernel abstraction,
we demonstrate how an application developer looking to work incrementally can
minimize changes to application code while gaining the portability benefits of
Kokkos.

\begin{lstlisting}[language=C++, caption={Example of converting a C++ dynamic array to a device View for incremental development, where SD is a struct containing XSBench simulation data.}, label={lst:kokkos}]
View<double*, LayoutLeft, HostSpace,
     MemoryTraits<Unmanaged>>
    u_concs(SD.concs, SD.length_concs);
SD.d_concs = new View<double*>("d_concs",
                               SD.length_concs);
deep_copy(*SD.d_concs, u_concs);
\end{lstlisting}

\subsection{Porting to RAJA}

In contrast to Kokkos, the RAJA portability ecosystem uses multiple libraries to
provide portability. Briefly, the RAJA library itself provides C++
lambda-capturing to allow developers to express portable computation. For memory
management, the developer can either write or use a custom portable memory
management library, or use the related Umpire~\cite{8907404} library, which
provides portable memory allocation primitives and memory pools. This separation
of concerns in the RAJA ecosystem provides facilitates incremental
porting of an existing codebase (i.e., portable compute first, then portable
data structures), avoiding more extensive refactoring.

In our case, we opt to take advantage of Umpire for CloverLeaf and XSBench,
which both have extensive existing code for managing and initializing data
structures. However, we encounter several challenges building the RAJA
applications. Relying on multiple independent libraries increases the expertise
required and frequency of errors in setting up build systems, a process that is
already complicated for a single library containing device kernels. Package
managers such as Spack~\cite{gamblin:sc15} can mitigate these problems for end
users, although this solution pushes the work of ensuring the
libraries build and install correctly onto the package maintainers.

\subsection{Porting to OpenACC}

OpenMP ports already exist for all applications, so creating similar OpenACC
ports where needed just requires a one-to-one conversion of the relevant OpenMP
pragmas to OpenACC. For example, \texttt{omp target teams distribute parallel
for} becomes \texttt{acc parallel loop}. This rote method makes our experience
with porting XSBench and CloverLeaf from OpenMP to OpenACC very productive. In
contrast to Kokkos and RAJA, working with existing data structures is highly
transparent in OpenACC, so long as the structures are plain old data (POD) and
do not contain pointers to CPU memory internally. In those more advanced cases,
which we do not encounter in this work, users must write more complex directives
to handle such data structures, convert them to simpler formats, or use
automatically managed memory if provided by the GPU
device~\cite{wolfe2018openacc}.

%% file: setup.tex
In this section, we describe the setup for the experiments conducted in this
work. We run all the applications on all five systems selected (listed in
Table~\ref{tab:systems}).

Table~\ref{tab:compilers} lists the compilers used with each programming model
alongside their versions. We use GCC 12.2.0 as the host compiler on NVIDIA
systems and ROCmCC 5.7.0 on AMD. We use CUDA version 12.2 on NVIDIA systems, and
HIP 5.7.0 on AMD systems, as well as Kokkos version 4.2.00 and RAJA v2023.06.1.
OpenACC, OpenMP, and SYCL all have different implementations provided by
multiple compilers on the systems where we perform our experiments. We test all
the available compilers for these models\footnote{For OpenMP we test Clang, GCC,
ROCmCC, NVHPC, CCE; for OpenACC we test Clacc, GCC, NVHPC; for SYCL we test
DPC++, AdaptiveCpp} and choose the best-performing compiler for each
application, model, and system. We perform this compiler-choice tuning to
reflect the fact that applications using these programming models will likely
test their code with all working compilers, and use in practice the
best-performing option.

\begin{table}[h]
  \centering
  \caption{Compilers and versions used for building each programming model implementation, by system type.}
{  \small %
  \begin{tabular}{lrr}
    \toprule
    Prog.~Model        & NVIDIA           & AMD \\ \midrule
    CUDA               & GCC 12.2.0       & N/A \\
    HIP                & N/A              & ROCmCC 5.7.0 \\
    SYCL*              & DPC++ 2024.01.20 & DPC++ 2024.01.20 \\
    Kokkos             & GCC 12.2.0       & ROCmCC 5.7.0 \\
    RAJA               & GCC 12.2.0       & ROCmCC 5.7.0 \\
    OpenMP$^{\dagger}$   & NVHPC 24.1       & LLVM 17.0.6 \\
    OpenACC            & NVHPC 24.1       & Clacc 2023-08-15 \\
    \bottomrule
  \end{tabular}
}
  \vspace{1mm}
  \flushleft
  \footnotesize{* We use AdaptiveCpp 23.10.0 for SYCL CloverLeaf.} \\
  \footnotesize{$^{\dagger}$ We use ROCmCC 5.7.0 for OpenMP su3\_bench on AMD GPUs.}
  \label{tab:compilers}
\end{table}

In all models except SYCL and OpenMP, the
best-performing compiler is consistent across applications on each system. For
the SYCL port of CloverLeaf, AdaptiveCpp is consistently superior, so we present
AdaptiveCpp results for that application and DPC++ for all others. For OpenMP,
ROCmCC wins on AMD systems for su3\_bench and Clang wins for all other
applications. Note also that we are unable to build CloverLeaf with Clacc due to
lack of support for the \verb|host_data| clause, and hence we cannot run
CloverLeaf on AMD systems with OpenACC.

We compile all proxy applications with `-O3' as well as fast math flags and
hardware specific instructions for approximate sqrt and division operations. For
AMD systems we also add `-munsafe-fp-atomics' as we found this to be broadly
beneficial to performance. Finally, for the Clacc compiler, we provide the flag
`-fopenacc-implicit-worker=vector-outer' at the recommendation of a developer,
as this flag will soon be enabled by default for Clacc.

We select input decks and command line inputs for each proxy application based
on recommended settings from their respective developers.
When given a choice of problem size, we select the largest representative
problem available that fits on all tested GPUs. We also choose the number of
iterations for each application to ensure about a minute of execution time,
so as to reduce variability. Section~\ref{sec:measurement} describes how we
modify the proxy applications to ensure consistent timings. We
present the final command line arguments in Table~\ref{tab:apps}.

\begin{table}[h]
    \centering
    \caption{Input parameters to the proxy applications.}
{    \small
    \begin{tabular}{lll}
    \toprule
    Application & Input parameters \\ \midrule
    BabelStream & \texttt{-n 1500 -w 150 -s \$((1<<29))}\\
    XSBench     & \texttt{-s large -m event -G hash -n 150 -w 15}\\
    CloverLeaf  & \texttt{--in clover\_bm64\_mid.in -w 52}\\
    su3\_bench  & \texttt{-l 32 -i 100000 -w 10000}\\
    miniBUDE    & \makecell[l]{\texttt{--deck bm2 -p 2 --wgsize 128 -i 10 }\\\texttt{--warmups 1}}\\
    \bottomrule
    \end{tabular}
}
    \label{tab:apps}
\end{table}

Note that for all cases tested the time spent in data movement is negligible
(less than 2\%) compared to time spent in device kernels, so our result figures
present \textbf{only} GPU kernel time. For all performance results presented we
run the application three times and present the average result. Variability is
low; the largest range of times recorded as a percentage of mean runtime for a
case is 3.3\%, and the mean is 0.1\%. We report total runtime for BabelStream
kernels rather than memory bandwidth in order to ensure that ``lower is better''
across all performance results we present. The values collected can be converted
to bandwidth (GB/s) by dividing the total data moved by the time.

\begin{figure*}[t]
    \centering
    \includegraphics[width=0.49\linewidth]{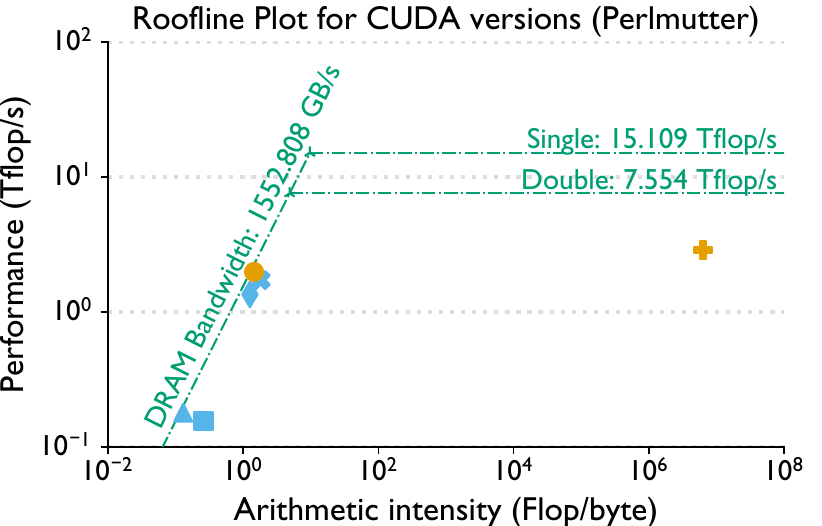}
    \includegraphics[width=0.49\linewidth]{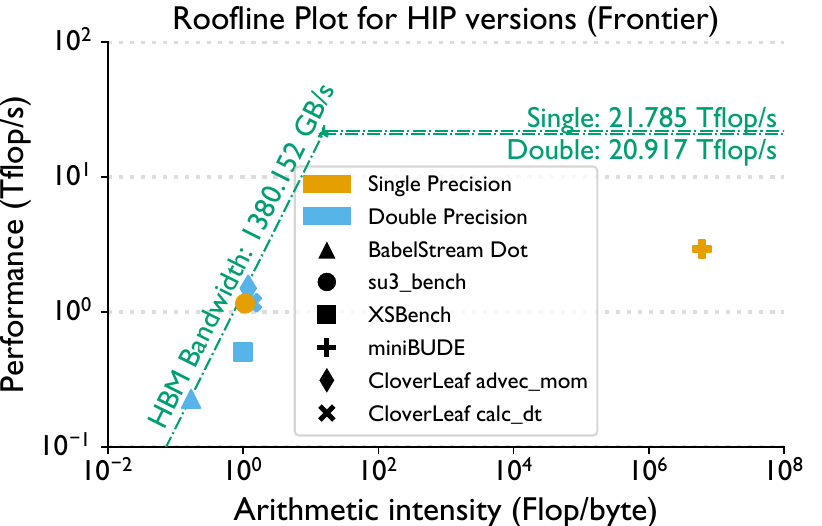}
    \caption{Roofline plots for the most time-consuming kernel in the CUDA (left)
      and HIP (right) versions of each application, from runs on Perlmutter (NVIDIA
      A100) and Frontier (AMD MI250X GCD) respectively. Orange points are single
      precision, and blue points are double precision. We plot each application's
      predominant precision.}
    \label{fig:roofline}
\end{figure*}

%% file: results.tex
We first present a roofline analysis of the native port implementations of each
application to understand their compute and memory behavior. Next, we present
the results of our model comparisons for individual applications and then
across all systems and applications.

\begin{table*}[t]
  \centering
  \caption{Key details of the major kernels in each proxy application used in
    the paper. CC is cyclomatic complexity and LN indicates the level of nesting
    in the kernel's main loop. Theoretical and achieved occupancy, shared memory
    per block, static instruction count, total DRAM traffic, and registers per
    thread are from NCU profiles of the CUDA implementations on Perlmutter.} {
    \small
  \begin{tabular}{llrrrrrrrr}
    \toprule
    Application & Kernel     & \makecell[r]{Cyclomatic\\complexity} & \makecell[r]{Main loop\\nesting\\level} & \makecell[r]{Grid, block\\sizes} & \makecell[r]{Theoretical,\\achieved\\occupancy} & \makecell[r]{Shared\\memory\\per block} & \makecell[r]{Total\\DRAM\\traffic} & \makecell[r]{Static\\instruction\\count} & \makecell[r]{Registers\\per thread} \\
    \midrule
    BabelStream & copy       & 1                                    & 1                                       & (524288, 1024)                   & 100\%, 81.8\%                                   & 0 B                                     & 8.6 GB                             & 12                                       & 16 \\
    BabelStream & dot        & 5                                    & 2                                       & (432, 1024)                      & 100\%, 99.9\%                                   & 8.2 KB                                  & 8.6 GB                             & 48                                       & 16 \\
    XSBench     & xs\_lookup & 39                                   & 3                                       & (66407, 256)                     & 50\%, 46.4\%                                    & 0 B                                     & 156 GB                             & 670                                      & 49 \\
    CloverLeaf  & advec\_mom & 4                                    & 2                                       & (230491, 256)                    & 62.5\%, 56.9\%                                  & 0 B                                     & 1.5 GB                             & 405                                      & 43 \\
    CloverLeaf  & calc\_dt   & 8                                    & 3                                       & (256, 256)                       & 62.5\%, 29.3\%                                  & 2.1 KB                                  & 4.6 GB                             & 643                                      & 47 \\
    su3\_bench  & k\_mat\_nn & 3                                    & 4                                       & (294912, 128)                    & 100\%, 89.7\%                                   & 0 B                                     & 624 MB                             & 59                                       & 26 \\
    miniBUDE    & fasten     & 29                                   & 4                                       & (256, 128)                       & 50\%, 14.5\%                                    & 0.7 KB                                  & 1.8 MB                             & 357                                      & 62 \\
    \bottomrule
  \end{tabular}
}
  \label{tab:kernels}
\end{table*}

\subsection{Roofline analysis}

Figure~\ref{fig:roofline} provides the empirical rooflines for the NVIDIA A100
GPU on Perlmutter and AMD MI250X GCD on Frontier. It also plots the positions of
the most time-consuming kernels in the CUDA and HIP implementations of the five
proxy applications. For BabelStream, this is the \verb|dot| kernel, and for
CloverLeaf, this is \verb|advec_mom|. {\tt advec\_cell} and {\tt PdV}
are also highly time-consuming, but have similar positions on the roofline. We
also plot \verb|calc_dt|, as it has the longest per-invocation execution time in
CloverLeaf. miniBUDE, XSBench, and su3\_bench contain a single computational
kernel each. We plot each kernel for the predominant floating-point precision
used. We observe that all kernels evaluated are memory-bound except for
miniBUDE, which is highly compute-bound, on both architectures. Among the
memory-bound apps, on both systems BabelStream \verb|dot| is the most
memory-bound (i.e., furthest to the left). This is expected given that
BabelStream is a memory bandwidth benchmark. CloverLeaf and su3\_bench are much
closer to the knee point on both systems, while XSBench has substantially
different arithmetic intensity on both systems --- 0.26 on Perlmutter, 1.00 on
Frontier. It is possible that XSBench heavily utilizes some instruction types
that are accounted differently between NVIDIA and AMD's counters used for
roofline plotting. Except for XSBench and miniBUDE, all of these kernels are
relatively close to the roofline, suggesting these CUDA and HIP versions are
relatively close to optimal for the algorithms they implement.

\begin{figure*}[t]
    \centering
    \includegraphics[width=\linewidth]{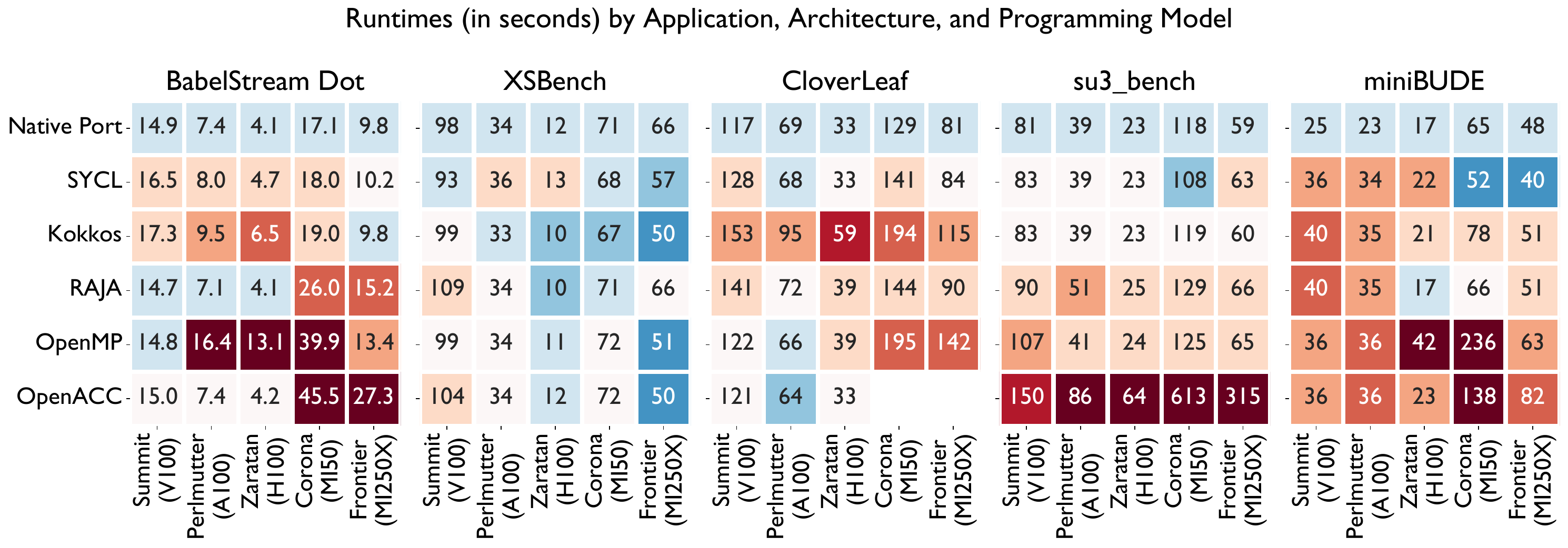}
    \rule{\textwidth}{0.2mm}
    \includegraphics[width=\linewidth]{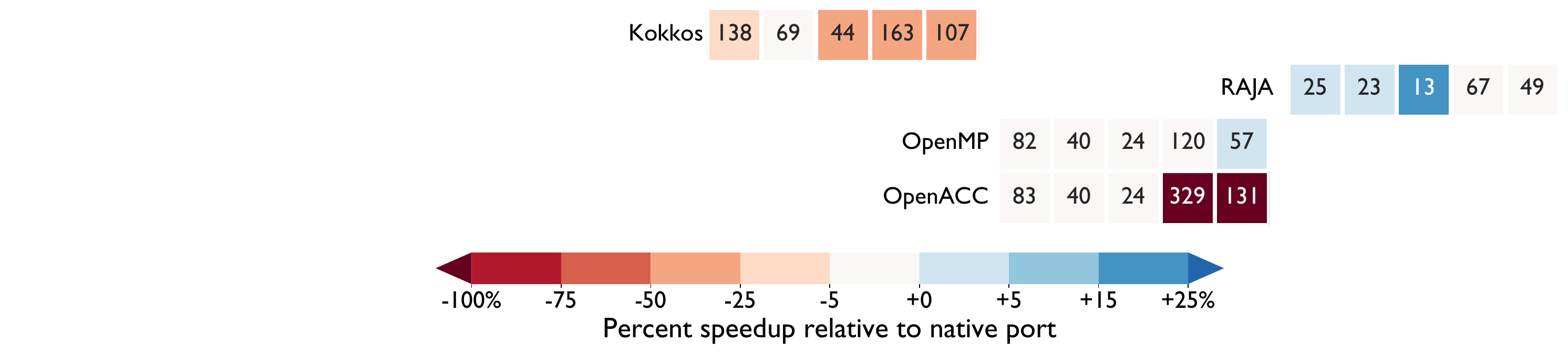}
    \caption{Execution time (indicated as raw numbers in each cell) over three
      trials of all proxy applications across all systems and programming models
      (lower is better). The color of each cell indicates performance
      improvement or degradation relative to the native port for that system and
      application (blue is faster and red is slower than the native port).
      Execution times for optimized implementations are provided below the
      horizontal line.}
    \label{fig:summary}
\end{figure*}

Table~\ref{tab:kernels} provides additional details about the kernels compared.
We provide cyclomatic complexity (CC) to reflect control flow complexity in the
kernels and the number of loops nested in the main loop to reflect
dimensionality of potential parallelism. The theoretical and achieved occupancy,
shared memory per block, static instruction count, total DRAM traffic, and
registers per thread are taken from Nsight Compute profiles of each kernel on
Perlmutter.

\observe{obs:roofline}{Most major kernels examined in this work are memory-bound
  with the exception of miniBUDE, which is strongly compute-bound.}

\subsection{Analysis of individual applications}

Next, we present performance results for BabelStream \verb|dot| and all four
mini-apps in Figure~\ref{fig:summary}. Each heatmap
cell represents the \textbf{total execution time} across all kernel invocations in each application.
Note that each value is the mean of three separate runs,
with the maximum difference between any run and this mean being
3.3\% (as described in Section~\ref{sec:method}). Also,
while we do measure data movement time, we do not report it here, as it is
consistently negligible ($<$2\% of runtime) compared to the time spent in the
GPU kernels. The ``Native Port'' row in each plot represents CUDA performance on
Summit and Perlmutter (the NVIDIA systems) and HIP performance on Corona and
Frontier (the AMD systems). We discuss observations derived from each
mini-app in turn.

\subsubsection{BabelStream}

\begin{figure}[h]
    \centering
    \includegraphics[width=\linewidth]{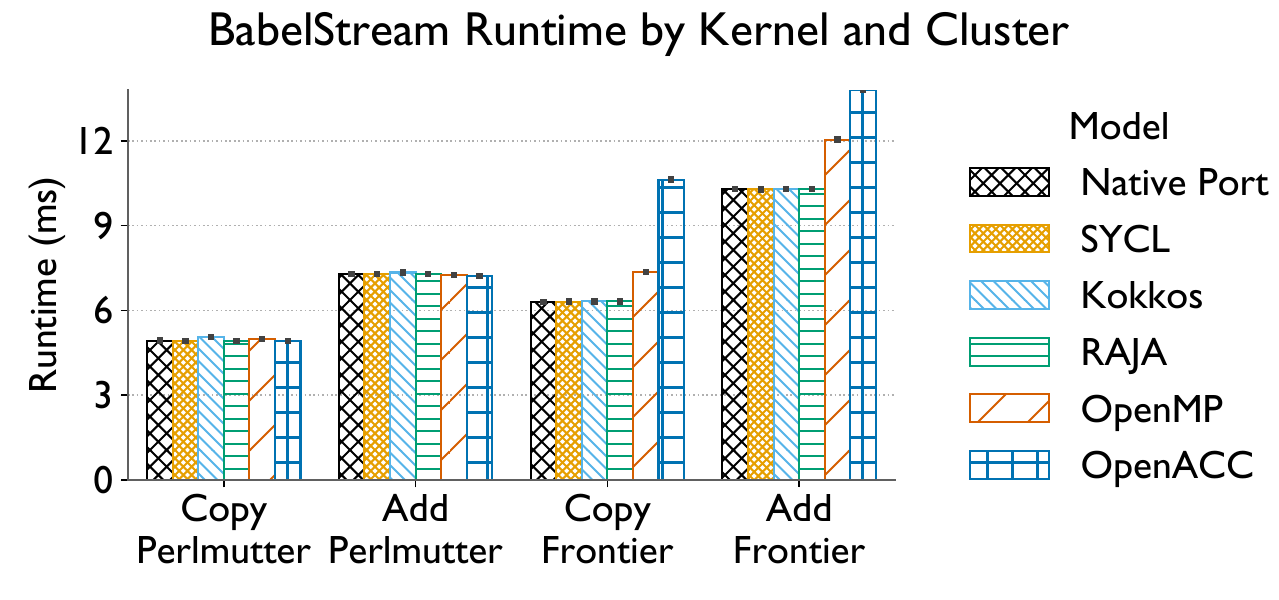}
    \caption{Execution time by model for Copy and Add kernels in
      BabelStream on Frontier (MI250X GCD) and Perlmutter (A100).}
    \label{fig:babelstream}
\end{figure}

BabelStream contains five kernels: \verb|copy|, \verb|add|, \verb|mul|, {\tt
  triad}, and \verb|dot|. All of these kernels are simple one-line memory-bound
operations. \verb|dot| is unique in that it utilizes a reduction operation to
compute a dot product, making it a useful simple benchmark of reduction
operations across programming models. We observe that for all kernels except
\verb|dot|, performance across programming models is highly consistent with the
native port. Figure~\ref{fig:babelstream} displays the performance of the
\verb|add| and \verb|copy| kernels across implementations on Frontier and
Perlmutter as a demonstration. OpenACC on AMD systems is the only significant
outlier, likely arising from overhead of the Clacc compiler translation
approach.

Performance in \verb|dot| is in contrast much more variable, which is why it is the only one reported in
Figure~\ref{fig:summary}. Usually, the portable programming models offer similar
or worse performance than the native port. Kokkos performs moderately worse than
the CUDA on Perlmutter and \zaratan{}, while RAJA performs moderately worse than
HIP on Corona and Frontier. Notably, OpenMP performs significantly worse
than the native port on Perlmutter, \zaratan{}, and Corona, and moderately worse
than HIP on Frontier. OpenACC performs significantly worse than HIP on AMD
systems.

In one notable exception, for RAJA BabelStream \verb|dot| on Perlmutter, we
observe that RAJA takes advantage of warp-level primitives and shared
memory to perform the reduction, maximizing utilization of hardware-specific
features for such operations. This allows RAJA to moderately out-perform
CUDA on all NVIDIA systems.

\observe{obs:babelstream:a}{Simple BabelStream kernels (other than {\tt
    dot}) are dominated by memory-bound operations, and all programming models
  can provide performance portability for them, except OpenACC on AMD.}

\observe{obs:babelstream:b}{For {\tt dot}, a reduction kernel, SYCL comes close to providing
  performance portability across all systems, and RAJA and OpenACC provide good performance
  on NVIDIA systems.
  OpenMP and OpenACC struggle to provide performance portability for {\tt dot}.}

\subsubsection{XSBench}

XSBench runs a single, long kernel which performs a large quantity of binary
searches. In this case, all programming models achieve near or moderately better
performance than the native port. In some cases, particularly on Frontier for
all models except RAJA, the portable programming models outperform HIP. Using
Omniperf to profile XSBench, we observe that the HIP port achieves lower Gflop/s
and lower L1 cache bandwidth, while Kokkos uses a larger workgroup size and
arranges L1 cache read requests in a larger number of smaller requests for a
similar number of bytes. This suggests Kokkos is selecting a more ideal
workgroup size and arranges data access patterns more efficiently for AMD GPUs
in XSBench. Meanwhile, OpenMP appears to take advantage of Local Data
Share (LDS) implicitly, reducing stalls for accesses to memory, while HIP does not.

XSBench is a performance test case used in the development of LLVM OpenMP
offloading, which Clacc also uses for OpenACC on Frontier, helping explain why
both directive-based models perform so well with XSBench. However, given that
Kokkos is a C++ abstraction over HIP code, it is surprising that it can
outperform HIP. We note that HIP XSBench performance on Frontier is only
slightly better than HIP XSBench on Corona, suggesting that the XSBench HIP
implementation is not a fully optimized and mature baseline.

Documentation for XSBench indicates that developers used the Hipify tool to
create the XSBench HIP port, and in comparing the HIP and CUDA versions it is
clear that they are identical aside from simple substitution of CUDA syntax for
HIP syntax. The XSBench kernel is also notably more cyclomatically complex and
longer than other kernels we examine (Table~\ref{tab:kernels}). Together, these
observations suggest that HIP kernels with more complex control flow translated
directly from CUDA without additional optimization may not guarantee optimal
performance on AMD GPUs, which have significantly smaller cache capacity per
thread workgroup relative to NVIDIA GPUs. More broadly, the case of XSBench
demonstrates how portable programming models are able to achieve matching or
even superior performance for more complex kernels with a similar level of
development effort as compared to vendor programming models.

\observe{obs:xsbench}{All programming models can achieve competitive performance
  portability for XSBench, a kernel with relatively complex control flow where
  the implementation style between portable and native ports is highly similar.}

\subsubsection{CloverLeaf}

As a larger proxy application with many kernels including structured stencil
operations as well as a reduction operation in \verb|calc_dt|, CloverLeaf
encompasses multiple types of GPU kernels. Nevertheless, several programming
models achieve broadly consistent performance compared to native ports on both
NVIDIA and AMD GPUs. SYCL in particular achieves slightly better performance
than CUDA on \zaratan{}. OpenACC slightly outperforms CUDA on both Perlmutter
and \zaratan{}, mostly due to improved performance in smaller kernels like
\verb|calc_dt|, but unfortunately we are unable to compile CloverLeaf with
OpenACC on AMD systems at this time due to lack of support for the
\verb|host_data| clause. RAJA performance is slightly worse than native ports
across systems.

Figure~\ref{fig:cloverleaf} breaks down CloverLeaf performance into major
constituent kernels to illustrate where performance differences arise for
outlier cases. First, CloverLeaf performance for OpenMP on Frontier is a
notable outlier. We find that compared to HIP the OpenMP port spends
significantly more time in the \verb|advec_*| and \verb|PdV| kernels. OpenMP
achieves less than half the L1 cache bandwidth in this kernel, as well as a
roughly 40\% lower L2 cache hit rate and 30\% higher rate of stalls on L2 cache
data, relative to HIP.

\begin{figure}[h]
    \centering
    \begin{subfigure}{0.204\textwidth}
      \centering
      \includegraphics[width=\linewidth]{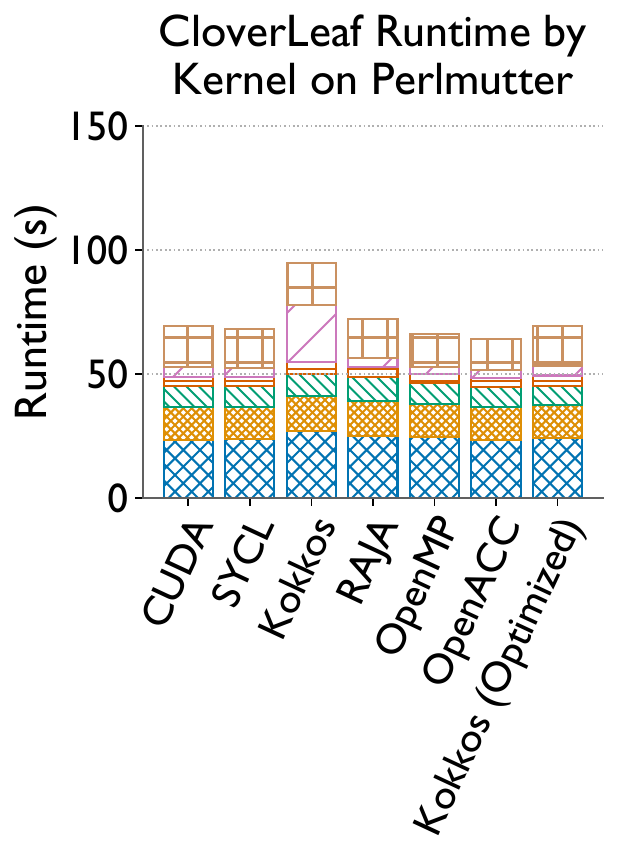} %
    \end{subfigure}%
    \begin{subfigure}{0.296\textwidth}
      \centering
      \includegraphics[width=\linewidth]{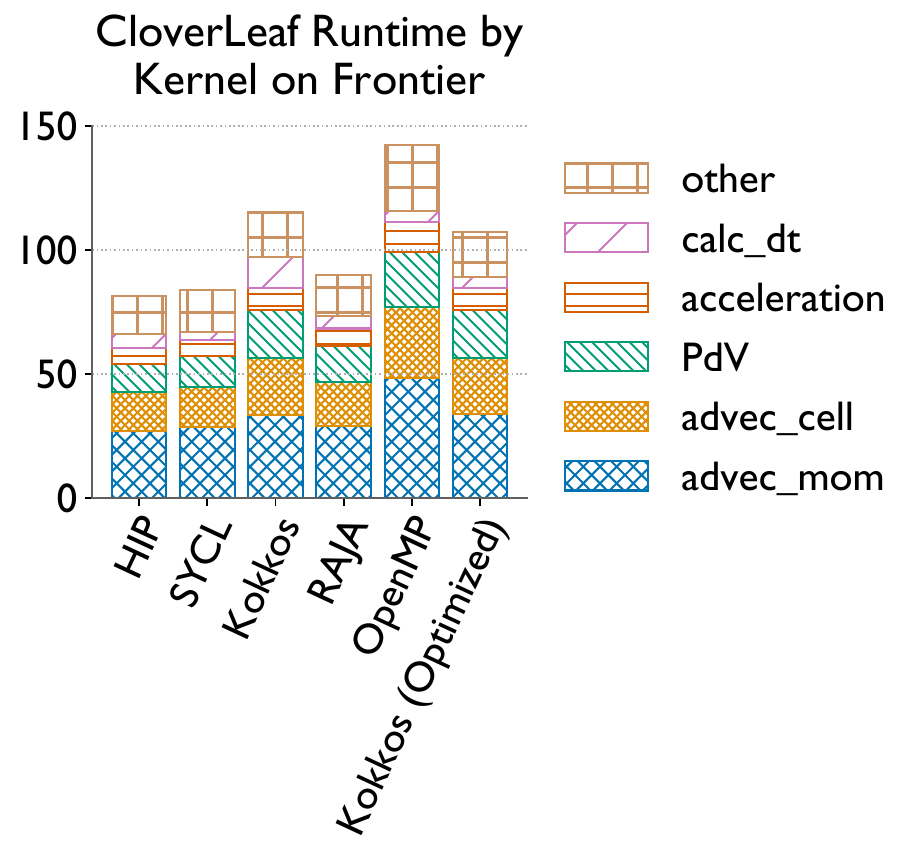} %
    \end{subfigure}
    \caption{Execution time by model broken down by kernel for CloverLeaf runs
      on Perlmutter (left, A100) and Frontier (right, MI250X GCD).}
    \label{fig:cloverleaf}
\end{figure}

Kokkos performance in CloverLeaf is also a notable exception. We observe that
the Kokkos port of CloverLeaf spends longer in the \verb|calc_dt| reduction
kernel relative to other ports, particularly on NVIDIA systems, like Perlmutter.
In Nsight Compute, we find that the Kokkos port
achieves fewer eligible warps on average, mostly due to barrier warp stalls,
which we do not observe in the other ports. Comparing the implementations of the
\verb|calc_dt| between ports, we find that Kokkos is the only one to use a 2D
reduction instead of collapsing the kernel into a 1D reduction. We adjust the
Kokkos port to use a 1D scheme, bringing Kokkos \verb|calc_dt| performance
closer to the native port on all studied systems, and no longer observe barrier
warp stalls in the new profile. As presented in Figure~\ref{fig:summary}, Kokkos
CloverLeaf performance improves on all systems with this change. The benefit is
greater on NVIDIA, where Kokkos's performance on the other three
significant kernels nears that of the native ports.

\observe{obs:cloverleaf}{For CloverLeaf's memory-bound kernels, which employ stencil
  operations, SYCL and RAJA can consistently provide performance
  portability and OpenMP, Kokkos and OpenACC struggle with providing portable performance.}

\subsubsection{su3\_bench}

su3\_bench is a single-kernel proxy application that computes a matrix
multiplication on complex numbers with a relatively deep nested loop hierarchy
(Table~\ref{tab:kernels}). Generally, performance across programming models is
fairly consistent, with the obvious exception of OpenACC before our
improvements.

The OpenACC port for su3\_bench in particular suffers from insufficient exposed
parallelism, even on NVIDIA GPUs. The su3\_bench OpenACC port originally
generates code with only 36 threads per block, despite iterations being assigned
to blocks of size 128. This leads to fewer active threads per block. We address
this issue by collapsing all four loops, exposing more parallelism.

We also find that both OpenMP and OpenACC generated twice as many global loads
and stores as CUDA, due to a misaligned complex number struct. OpenMP, OpenACC,
and SYCL do not provide a GPU-native complex type. We declare this struct
aligned to {\tt sizeof(T) * 2}, resulting in a single load and store for each
complex number in the array. On AMD this optimization has no effect. As
presented in Figure~\ref{fig:summary}, OpenACC benefits strongly from this
combination of optimizations, whereas OpenMP achieves modest speedups.

For RAJA, we observe substantially lower arithmetic intensity in L1 and L2 cache
compared to native ports, suggesting the RAJA port loads unnecessary data from
memory more often, although the overall impact on performance portability of
this limitation is relatively low.

\observe{obs:su3bench}{For su3\_bench, a memory-bound kernel with a deep loop
  nest, almost all programming models can achieve approximate performance
  portability as long as sufficient parallelism is exposed and struct
  declarations are aligned, with the moderate exception of RAJA and stronger
  exception of OpenACC.}

\subsubsection{miniBUDE}

miniBUDE, a single-kernel app, is by the the most challenging proxy application
for the programming models we test. miniBUDE is unusual compared to our other
proxy applications in that it is highly compute-bound, and leverages thread
coarsening to hide memory latencies using instruction-level parallelism within
the kernel. It also exerts the highest register pressure in the native CUDA
implementation relative to other kernels we examine (Table~\ref{tab:kernels}).
No programming model is able to achieve consistent performance with the native
port, except for RAJA after our improvements. We note that SYCL does achieve
superior performance compared to HIP on AMD devices.

In comparing the Kokkos, RAJA, SYCL, and CUDA versions of miniBUDE, we notice
that the RAJA version is not making use of shared memory, while the Kokkos,
SYCL, and CUDA ports are. RAJA recently added features for dynamically
allocating shared memory inside a kernel, a feature needed in miniBUDE since the
forcefield data is input-dependent in size, so we modify RAJA miniBUDE to use
shared memory for this data.

This optimization improves RAJA performance on NVIDIA systems, with little
impact on AMD, leading to an overall increase in portability (see
Figure~\ref{fig:summary}). After the change RAJA performance comes very close to the
CUDA performance on Perlmutter and 27\% faster than CUDA on \zaratan{}, an
impressive gain since other models already using shared memory do not get this
close on NVIDIA systems. At the time of writing we are unable to add dynamic
shared memory allocation inside the kernel for the OpenMP and OpenACC ports due
to lack of support.

The OpenMP port of miniBUDE appears to allocate an order of magnitude more Local
Data Share (LDS) bytes than HIP does, limiting the number of active compute
units and thus reducing the degree to which memory access latency can be hidden.
Comparing the OpenMP port to CUDA, we observe a significant increase in
registers used per thread (86 vs. 62) and dramatically more static instructions
(1463 vs. 357). Other models encounter similar, but less pronounced, issues: for
example, Kokkos uses 69 registers per thread and generates a 797-instruction
kernel. For both Kokkos and OpenMP these overheads correspond to a 500\%
increase in cycles spent in L2 cache activity as well as ~50\% increases in DRAM
and L1 cache cycles.

\observe{obs:minibude}{For miniBUDE, a highly compute-bound kernel relying on
  shared memory, RAJA (after adding shared memory support) can provide very
  competitive performance portability where most other models, especially OpenMP
  and OpenACC, struggle due to register pressure and increased generated
  instruction counts. SYCL is also highly competitive, but only on AMD systems.}

\subsection{Evaluating performance portability across applications after optimizations}
\input{results-models}

%% file: results-models.tex
From analysis of the lower portion of Figure~\ref{fig:summary}, containing
results after our optimizations, we can make several general observations about
the performance portability enabled by each programming model.

\subsubsection{Language extensions}

CUDA is the best or within 3\% of the best performing model in eleven out of
fifteen cases. For these applications, this is a useful validation of the
maturity of the CUDA baseline for each application, and confirms our expectation
that the low-level vendor model would be the most performant and portable across
GPUs from that vendor.

Meanwhile, for most cases on AMD systems, including CloverLeaf, BabelStream {\tt
  dot}, and su3\_bench on Frontier, AMD's HIP programming model achieves the
best performance, as expected. However, in multiple instances, HIP does not
achieve the best performance, particularly for XSBench, as discussed under that
application.

Finally, SYCL performs better than HIP in five out of ten cases on AMD systems.
As a lower-level language extension, similar to CUDA or HIP, this is not
necessarily surprising. In some cases, SYCL is able to improve on CUDA or HIP
performance, and even where SYCL is more than 3\% slower than a native port, it
is never the worst-performing port except in XSBench on Perlmutter and
\zaratan{}, where it is is only 5.3\% and 10\% slower, respectively. SYCL is the
fastest non-native programming model in more cases than any other model, at
eleven out of twenty-five total application and system pairs, and six of these
are on AMD systems.

\observe{obs:cudahip}{On NVIDIA systems, CUDA almost always performs at or near
  the best observed performance, whereas on AMD systems there are some cases,
  in particular for XSBench, where other models are significantly faster than HIP.}

\observe{obs:sycl}{SYCL performance is often competitive with CUDA and HIP, and relatively
  stable across system and application pairs, with the exception of miniBUDE on
  NVIDIA GPUs.}

\subsubsection{C++ abstraction libraries}

Kokkos and RAJA compare favorably with CUDA and HIP on NVIDIA and AMD systems,
with one of the two ports either nearing or exceeding the native port's
performance on every combination of system and app, besides those involving
CloverLeaf on any system or miniBUDE on Summit and Perlmutter. While which model
is more performant is very application-dependent, we can observe that RAJA tends
to perform more competitively for NVIDIA systems, and Kokkos tends to have an
advantage on AMD systems.

\observe{obs:kokkosraja}{Kokkos and RAJA are competitive with CUDA and HIP on
  many system and application pairs, with a slight preference for RAJA on NVIDIA
  GPUs and Kokkos on AMD GPUs.}

\subsubsection{Directive-based models}

OpenMP performance can be slower than the native baseline, achieving
significantly better performance than the baseline only for XSBench on Frontier.
OpenMP is able to achieve rough parity with the native baseline in twelve out of
twenty-five cases.

On NVIDIA systems, OpenACC generally achieves more consistent performance with
the baseline, but is often worse than OpenMP and further worse than HIP on AMD
systems, likely because it is employing the same LLVM OpenMP offloading runtime
through the Clacc compiler. Per Clacc developers, there is some overhead due to
suboptimal translation of OpenACC to OpenMP within Clacc which will be addressed
in a future release.

\observe{obs:openmpopenacc}{OpenMP is slower than other implementations in
  roughly half our cases, and OpenACC struggles with AMD systems.}

\subsection{Performance portability metric evaluation}

\begin{figure}[h]
    \centering \includegraphics[width=\linewidth]{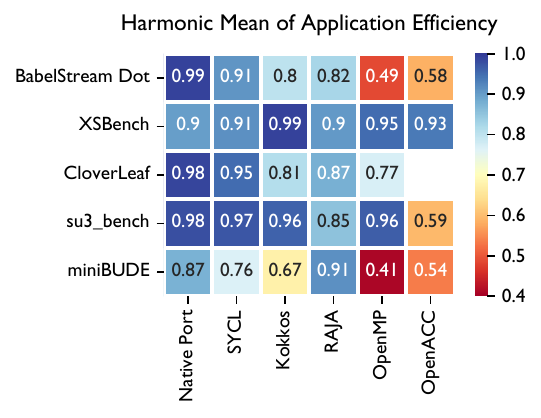} \caption{\ppm
    of GPU kernel performance for each programming model and application
    combination, after optimizations. Applications are listed in ascending order
    of arithmetic intensity. Note for OpenACC we are unable
    to compile CloverLeaf on AMD systems.} \label{fig:ppm}
\end{figure}

Figure~\ref{fig:ppm} displays the \ppm metric for each programming model and
proxy application combination after applying the optimizations described above.
The ``Native Port'' column provides context, indicating what the metric would
report if a team decided to maintain both a HIP and CUDA version of the
application. We are unable to run CloverLeaf with OpenACC on AMD systems, so
that cell is zero per the official formulation of the metric\footnote{When only
considering NVIDIA systems, the value is 0.98.}. According to \ppm, we observe a
moderate preference for SYCL, RAJA, and Kokkos among the portable programming
models, in roughly that order, and for OpenMP over OpenACC within
directive-based models.\footnote{We also compared \ppm to
$\overline{\pp}$~\cite{marowka2023comparison}, which uses the arithmetic mean
instead of the harmonic mean (results not shown). This penalizes low outliers
much less, but the overall ordering of results remains the same with both
metrics.}

\observe{obs:ppm}{Summarizing our \ppm results, we find that SYCL most
consistently achieves performance portability for our tests, followed closely
by RAJA and Kokkos.}

%% file: conc.tex
In this paper, we empirically evaluated seven GPU programming models and
directly compared their capabilities for enabling performance portability. We
performed this evaluation on some of the fastest supercomputers in the world
using proxy application codes that represent real scientific workloads. We
developed a Spack-based methodology to substantially lower the barrier for
future experiments comparing portable programming models. We invested
significant effort in ensuring each proxy application's implementations in each
model can be easily built and run on additional systems, and we plan to
open-source these efforts, sharing them with the broader HPC community. Overall,
compared to prior comparative
studies~\cite{deakin2019performance,deakin2020tracking} we find improved
performance portability across models, particularly for SYCL.

After our optimizations, a few broad outliers remain in the performance results
we studied which may be of interest to developers looking to
choose a programming model. We highlight the frequent gap between OpenACC and
OpenMP performance on AMD systems, generally poor reduction performance in
OpenACC and OpenMP, poor reductions on AMD systems with RAJA, and consistent
difficulty with the compute-bound and register-intensive miniBUDE for all
programming models and systems. For application, compiler, and programming model
developers, we present several insights from our experiences as well as
suggestions for future investment of effort towards performance portability:
\begin{itemize}
\item Successfully building all of these applications across systems is not
  trivial, especially for a multi-library portability suite like RAJA.
  Additional robustness in -- and documentation for -- this build process may
  enable app developers to more easily test competing programming models.
\item Our ability to identify bottlenecks depended heavily on profiling tools.
  Improving the quality of these tools for new programming models and hardware
  architectures will be critical to enabling performance portability. Line-level
  stall attribution is a crucial capability missing from Omniperf at the time
  of writing.
\item Reduction operations continue to be a major bottleneck, as observed in
  prior studies, and work on improving compiler handling of reductions would
  close some of the major remaining performance gaps between portable models
  and native baselines.
\item The example of miniBUDE demonstrates that performance in a compute-bound
  kernel with high register pressure can be highly sensitive to the choice of
  a portable programming model. Identifying techniques to reduce spilling of
  registers to memory and instruction count bloat when adopting a programming
  abstraction may help users maximize arithmetic bandwidth.
\item The ability to separate correctness and performance concerns in these
  models was critical in identifying the optimizations we describe, as it
  allowed us to tune ports without invalidating scientific results. Exposing and
  documenting more semantic-preserving performance \say{knobs} within each
  model may provide developers with a wider range of options to improve the
  performance portability of their applications.
\end{itemize}

%% file: ack.tex
This material is based upon work supported in part by the National Science
Foundation (NSF) under Grant No.~2047120, the NSF Graduate Research Fellowship
Program under Grant No.~DGE~2236417, and the U.S.~Department of Energy (DOE),
Office of Science, Office of Advanced Scientific Computing Research, DOE
Computational Science Graduate Fellowship under Award No.~DE-SC0021. This work
was performed in part under the auspices of the U.S.~DOE by Lawrence Livermore
National Laboratory under Contract DE-AC52-07NA27344 (LLNL-CONF-855581).

This research used resources of the Oak Ridge Leadership Computing Facility at
the Oak Ridge National Laboratory, which is supported by the Office of Science
of the U.S.~DOE under Contract No.~DE-AC05-00OR22725, and of the National
Energy Research Scientific Computing Center (NERSC), a U.S.~DOE Office of
Science User Facility located at Lawrence Berkeley National Laboratory,
operated under Contract No.~DE-AC02-05CH11231 using NERSC awards
DDR-ERCAP0025593 and DDR-ERCAP0029890.